\documentstyle[prd,aps]{revtex}
\tighten
\begin{document}
\def\breaksto{\mathop{\kern0pt \to}}
\draft
\twocolumn[\hsize\textwidth\columnwidth\hsize\csname
@twocolumnfalse\endcsname \title{COSMIC STRINGS ARE CURRENT-CARRYING.}
\author{Anne-Christine DAVIS$^{1,2}$ and Patrick PETER$^{2,3}$}
\address{$^1$CERN - Theory division, CH-1211 Geneva 23, Switzerland,\\
$^2$D.A.M.T.P., Silver Street, Cambridge CB3 9EW, England,\\
$^3$D.A.R.C. (UPR176, CNRS), Observatoire de Paris, 92195 Meudon,
France.}
\date{June 27, 1995 \ \ -- \ \ CERN-TH/95-173, DAMTP-95-34,
HEP-PH/9506433}
\maketitle \begin{abstract}
  A  synthesis of previous work done   on the microscopic structure of
  cosmic strings in realistic models  is made and reveals that strings
  are expected to be not only  superconducting in the sense of Witten,
  but also generically current-carrying, either at the GUT scale or at
  the electroweak scale.  This applies to any GUT string forming model
  leading to the  standard electroweak theory  as a  low energy limit.
  The current  consists of  charged  vector bosons.  Cosmological
  consequences are briefly discussed.
\end{abstract}
\pacs{PACS numbers: 98.80.Cq, 11.27+d}
\vskip2pc]

\section{Introduction}

Most cosmological applications of cosmic strings~\cite{kibble,vil}
have  been essentially  based on noncurrent-carrying  vortices, namely
those having no internal structure  and hence describable by
means of  the simple Goto-Nambu action.   They  would have appeared as
topological defects   during  a very early  phase  transition  such as
predicted by  many  Grand Unified  Theories  (GUT). They have been
shown to  form a network that  could explain the large scale structure
of   the universe~\cite{vil} and  the  temperature fluctuations in the
cosmic microwave  background  radiation~\cite{vil}.  Because  of their
origin  in particle  physics models and   their high predictive  power
(only one free parameter, namely the energy per unit length $U$), they
represent the major   opponent to  the  almost standard   inflationary
scenario~\cite{infla}.

Cosmic  strings  of  the  current-carrying type, such  as  proposed by
Witten~\cite{witten},  have recently received  a revival of interest in
particular  thanks   to their  potential usefulness  in  producing the
observed  baryon number  asymmetry~\cite{baryons}. Following
this  original  attempt~\cite{witten} to show that currents
could appear along string's  worldsheets, a number of particle
physics   models were shown  (actually designed  for  that purpose) to
similarly  exhibit strings  endowed  with superconducting  properties.
Hence the question of how generic this feature may be.  The purpose of
this  letter is to address this  question. We show that within the
framework of the standard GUT paradigm with the  electroweak model as a
low energy limit,  it is inconceivable  that a phase transition in the
early  universe produced strings that  would still be structureless, as
the ones mostly used in any numerical simulation.

Currents in strings change our understanding of the string
cosmology in particular because the scaling property of the network might
well be absent, or modified, in such models. Thus, it sets various
constraints   on  the underlying  particle  physics  models  on scales
otherwise inaccessible. In fact, as we shall see later, because of the
so-called vorton problem~\cite{vortons}, there are already cases which
are  ruled out  even  though only very little  is  currently known  about
current-carrying cosmic string  cosmology.  Such  a strong consequence
makes it clear   that this new branch  of  cosmology has been  largely
underestimated until now and deserves much more attention.
Among many other  possibilities, let us  just mention
for instance  here the obvious   one that primordial  magnetic  fields
might be generated. This could fairly well lead to a complete revision
of our views concerning large scale structure formation.

The reason  why  currents must  be  expected to be  present  in cosmic
strings is  twofold.  First, the existence  of  a symmetry restoration
region~\cite{symres}    and  then    that    of   spontaneous  current
generation~\cite{scg}; both are generic  phenomena in particle physics
models predicting  cosmic strings.  As we  shall see, these mechanisms
are responsible for  the appearance of  a current first, and  then for
its stability. This yields three distinct possibilities for the energy
scales     involved,   namely,   using       the     notations      of
Ref.~\cite{supersonic}, the  mass scale $m$ at  which the string forms
(i.e.,  with $U\sim m^2$)  and  $m_\star$ being characteristic of  the
current intensity. The careful analysis that follows reveals that only
a  few numerical  values for  these string's  parameters are  actually
realizable in Nature, namely  $m\sim m_\star\sim \eta_{_{\rm GUT}}\sim
10^{15}$~GeV the scale of Grand Unification, $m\sim \eta_{_{\rm GUT}}$
and $m_\star\sim\eta_{_{\rm EW}}\sim 100$~GeV the scale of electroweak
symmetry  breaking, or  an   intermediate scale  for  $m$ provided  it
exceeds $\eta_{_{\rm Int}}\sim 10^{10}$~GeV, where the bound is set up
by vorton formation, still with $m_\star\sim\eta_{_{\rm EW}}$, or
finally  similar scales $m\sim m_\star$,  given then~\cite{vortons} by
$\eta _{_{V}} \sim 10$~TeV.

This  work is organized as follows:  in section 2,  we set the problem
and the  notation; in section 3,  we summarize  the basics of symmetry
restoration around a cosmic  string, then section  4 is devoted to the
spontaneous  current  generation mechanism  while  section 5 uses both
effects to yield the conclusion that cosmic strings are expected to be
current-carrying  independently   of the  underlying  particle physics
model.

Finally, we briefly  discuss cosmological consequences including  the
corresponding constraints due for instance to the vorton problem.

\section{The problem}

Let us first fix the notation that is used throughout. For the sake of
generality, we shall consider a  theory, effective or actual, with the
following scheme of symmetry breaking:

$$G \breaksto^\Phi H \breaksto^h \cdots \to SU(3)\times
SU(2) \times U(1) \hskip2cm $$
\begin{equation} \hskip4cm \breaksto^H SU(3)\times U(1),
\label{scheme}\end{equation}
where $G$ might already be the result of previous symmetry breaking(s)
(hence it  is not necessarily simple),  and $\Phi$,  $h$ and $H$ stand
for  the various  Higgs fields  responsible  for the phase transitions
(arrows) over which they are symbolized on Eq.~(\ref{scheme}), meaning
for instance that $H$ is the ordinary  $SU(2)$ doublet of the standard
electroweak model. Note also that nothing prevents $h$ to be identical
to either $H$ or $\Phi$ and what we  clearly request in fact is solely
that there are at least two symmetry breakings,  one of them being the
electroweak   one.  Moreover, we   shall assume $\pi _1 [G/SU(3)\times
U(1)]\not\sim \{ 0\}$ so that strings are formed at some stage.

Let us  now turn  more specifically  to the string  forming model. The
Lagrangian  density we  are  interested in    is the GUT  one  without
fermions, namely
$$ {\cal L} = {1\over 2} (D_\mu \Phi )^\dagger (D^\mu
\Phi ) + {1\over 2} (D_\mu h )^\dagger (D^\mu h ) \hskip2cm $$
\begin{equation} \hskip4cm - {1\over 4}
F^a_{\mu\nu} F^{a\mu\nu} - V(\Phi ,h),\label{lag}\end{equation}
where
\begin{equation} D_\mu \Phi \equiv (\partial _\mu +ig T_a A^a _\mu )
\Phi , \  \ \ D_\mu  h\equiv (\partial _\mu + ig  \tau_a A^a_\mu ) h
,\label{Dmus}\end{equation}
are the covariant derivatives of the  Higgs fields $\Phi$ and $h$, the
former transforming under $G$ according to the representation given by
the $T_a$'s, basis for  the Lie algebra ${\cal  L}(G)$, and the latter
according   to the representation given  by   the $\tau_a$'s [note  in
particular that $h$ transforms trivially under  most of the generators
of  $G$ except for those that  span the subalgebra ${\cal L}(H)$]. The
kinetic  term   for the   gauge   vectors $A_\mu^a$  follows from  the
definition
\begin{equation} F_{\mu\nu}^a\equiv \partial_\mu A_\nu^a -
\partial_\nu A_\mu^a - g f^{abc} A_\mu^b A^c_\nu ,\label{Fmunu}
\end{equation}
where $g$  is  the  (potentially running)  coupling strength   and the
$f^{abc}$'s are the structure constants of the gauge group $G$ with
\begin{equation} [T_a, T_b] = if^{abc} T_c .\label{group}\end{equation}
  Note  also that  we  are using   a  metric with signature $+2$  when
  writing the Lagrangian~(\ref{lag}).

  A  vortex  solution  to  the Euler-Lagrange  equations  derived from
  Eq.~(\ref{lag}) is given classically by~\cite{alford}
\begin{equation} \Phi = f(r) \exp (-in\theta T_s )\Phi_0 ,
\label{phisol}\end{equation}
\begin{equation} T_a A^a_\mu = -\displaystyle{n\over gr} v(r) \delta
  _{\mu\theta}     T_s ,\label{gaugesol}\end{equation}
for  a  string  aligned  along the    $z-$axis in  rectangular   polar
coordinates.    $T_s$  is  the string's  generator    and the boundary
conditions are
$$f(0)=v(0)=0, \  \ \ \lim_{r\to\infty}  f(r) = \lim_{r\to\infty} v(r)
=1,$$ $\Phi_0$ being the  actual Higgs field's constant  configuration
that  minimises the potential  at  infinity. The scale  is set  by its
vacuum expectation  value  (VEV) $|\Phi  _0|^2  =  m^2$  following the
notation of the introduction.

To end  this section, and before  turning our attention to  the actual
symmetry restoration mechanism, let us consider the various terms that
generically will   be  present in   the  Lagrangian~(\ref{lag}).    In
particular, there should  be direct coupling  terms of the form $|\Phi
|^2 |h|^2$, given that  they are renormalisable.   Indeed, such a term
must be present, would  it be only at the  one-loop level, because the
number of broken  generators dim($G)-$~dim($H$) is,  as can be checked
by direct evaluation  of all possible  cases of Lie algebras and their
maximal subalgebras~\cite{algebras}, far greater than the rank of $G$.
Thus,  the  gauge bosons coupled  to  $\Phi$, in any gauge,  call them
$A_\mu^{(\phi  )}$, must  be coupled to   some gauge bosons coupled to
$h$, $A_\mu^{(h )}$ say.  Therefore, $\exists  T^{(\phi )}\in {\cal L}
(G)$ and $\exists T^{(h )}\in {\cal L} (H) ; [T^{(\phi)},T^{(h)}]\not=
0$.  So the corresponding structure constants $f^{(\phi ) (h)(h)}$ and
$f^{(h)(\phi)(\phi)}$  do not all    vanish and the  coupling  exists.
Similarily, it can  be shown that the  other interesting term coupling
$h$ and the string's  generator cannot be   made to vanish;  thus, $h$
carries $T_s$ hypercharge.   We shall  call  $X_\mu$ the gauge  vector
boson associated with $T_s$.

Both  the two terms  discussed in the  previous paragraph are required
for a charged-coupled  current to  exist in  the string's  core as  we
shall  now see.  Since we have  just shown  that  they are generically
present, this achieves our proof of  the current-carrying abilities of
cosmic-strings.

\section{Symmetry restoration}

A generic feature of cosmic strings formed at a scale $m$ is that they
lead to  restoration of lower energy symmetries.   For example,  a GUT
scale string restores  the electroweak symmetry  around it in a region
proportional to the electroweak scale~\cite{symres}. Obviously in this
region the electroweak particles are massless. Similarly, if there are
a series  of phase transitions, the  GUT string will restore these out
to a region   depending on the scale  of  the phase  transition. To be
concrete, let  us  consider a string   with gauge field  $X_\mu$.  The
string's field couples  to the Higgs  field $h$ of the  less symmetric
sector as we previously discussed,
\begin{equation} D_\mu h = (\partial_\mu - {i\over 2} \tilde g
\vec \tau \cdot \vec W_\mu -{i\over 2}\alpha X_\mu )h,
\label{DmuPhi}\end{equation}
where the coupling strength  $\alpha$ can either  be a direct coupling
-- as  expected in GUTs  --or induced by  loop corrections and we have
specifically    extracted  out the  part  depending    on the string's
generator $X_\mu$ from  the other ones which  we called $\vec  W_\mu$.
(Recall that $\alpha$ and $\tilde  g$ have in general  no reason to be
identical to $g$ unless the $A_\mu^a$'s  representation choice is set up
for this to  occur.)   In either   case, the term~(\ref{DmuPhi})   is,
again,    generic.   Putting   in    the    Nielsen-Olesen   form  for
$X_\mu$~\cite{Nielsen-Olesen}  for  a  unit  winding number string  as
given in Eq.~(\ref{gaugesol}), one solves the  equations of motion for
the  electroweak theory using  the ansatz~\cite{symres}$h = {\rm e}^{i
  \ell \theta} \eta  (r)$.  In particular,  the presence of the  extra
term in Eq.~(\ref{DmuPhi})  changes the $h$  and part of  the $\vec W$
equations   of motion --  in    the background~(\ref{gaugesol}) it  is
energetically   favourable that   there is   a  part of  $\vec W$  that
vanishes.

Calling $Z$ (in analogy with the electroweak model~\cite{symres}) that
part of $\vec W$ which  does not vanish,  one finds that in order  to
ensure finite  energy   at large   distances, $h$  and  $Z$ must  have
profiles  similar to  that of the Nielsen-Olesen~\cite{Nielsen-Olesen}
profile.  Thus, using a trial solution~\cite{symres}
\begin{equation} h = {m_\star \over 2} \left[ \matrix{
(r/  r_1)^{a/2} &  r  < r_1  \cr 1 &  r >  r_1 \cr} \right.  ,
\label{phi}\end{equation}
\begin{equation} GZ=GZ_\theta = -a \left[ \matrix{r / r_1^2 & r<r_1 \cr
1/r  & r> r_1 \cr }\right.  ,\label{gZ}\end{equation}
where $G$ is an effective coupling constant that can be calculated out
of the choice of $\tau_a$'s  (for instance in the  case of $h$  being
identical  to $H$, the    Higgs  doublet of  the  electroweak   model,
$G^2=g^2+g'^2$,   with   $g$   and   $g'$  the   gauge    couplings of
$SU(2)_{_{L}}$ and  $U(1)_{_{Y}}$), $m_\star$ is  the VEV of the Higgs
field $h$, and
$$ a = 2\ell + \displaystyle{\alpha \over g} \Theta (r-m^{-1})$$
with $\Theta$ the Heaviside step function.

Minimizing the energy integral, one  finds  that $m_\star r_1 \sim  1$
and the energy of the trial solution is $a m_\star^2$.  Thus, there is
a region of  $H$ symmetry restoration  around the $G$ scale  string of
order the $H$ scale.  We also note that if  $\alpha$ is large, then it
is energetically favorable for the Higgs field $h$  to wind around the
string,  i.e.,  for  $\ell\not=  0$.   The previous  discussion indeed
applies   to any intermediate  scale  that   would  be restored  in an
anologous fashion   because nothing of  what  has just been  stated is
explicitely dependent  on  the symmetry  breaking scheme and  the Higgs
structure. In particular, it can be applied (and it is in fact in this
context that  this mechanism  was originally derived~\cite{symres}) to
the case of a GUT string coupled to the electroweak sector, i.e., with
$h$ identified with  the doublet $H$ and  $m_\star  \sim M_{_{W}} \sim
100$~GeV.

This point being  settled, let  us  now turn  to the second  necessary
ingredient.

\section{Spontaneous current generation}

The  reason why cosmic strings are  generically superconducting and in
fact even current-carrying   is because  of the spontaneous   current
generation mechanism, which proceeds as follows.  Whenever the mass of
a  charged (or hypercharged,   provided  the  corresponding  generator
remains massless) vector boson gets lowered in some finite size region
surrounding    a   string's  core (e.g.,    the   region  of  symmetry
restoration), then it becomes energetically favourable for this charged
boson to acquire   a nonvanishing VEV   in the string's core,  thereby
effectively  breaking spontaneously   the  Lorentz symmetry  along the
string's worldsheet. Thus, one just has to prove  that in general such
a  charged boson exists  to prove  generic superconductivity of cosmic
strings.  The three distinct  regimes mentioned earlier are  then made
apparent once a close  examination of the various cases  is made.  Let
us  first  discuss  in  more detail   the current generation mechanism
itself.

Suppose the  Higgs field $h$  to be coupled to  a massless-hypercharge
(like the actual electric charge) carrying vector boson $C_\mu$ in the
sense  that at least  part (but possibly  all)  of its mass originates
from  $h$'s  VEV.  Then close  to  the string's core,  $C_\mu$ is less
massive than far from it because  of the symmetry restoration. Now the
field equations for  $C_\mu$ near the vortex,  where  $\Phi = h  = 0$,
read as a part of the general $A_\mu^a$ field equations
\begin{equation} \nabla ^\mu F^a_{\mu\nu} - g f^{abc} A^{b\mu}
F^c_{\mu\nu} =0 ,\label{Amueq}\end{equation}
which is  usually  solved in  vacuum by the  trivial solution $A^a_\mu
=0$, and so $C_\mu =0$. However, this solution is  nothing but a gauge
choice and it is possible to choose anything else without, at least in
vacuum,  changing the total  energy of the  configuration.  But we are
near a  string where  gradients of scalar   fields are important since
$\Phi  =0$  at $r=0$ is    topologically required  while bounding  the
overall energy dynamically   requires  $|\Phi |^2   =   m^2$ at  $r\to
\infty$.  As a result,  changing the solution of Eq.~(\ref{Amueq})  is
not  just a  gauge choice   and the  configuration having  the minimum
energy is one having nonzero $C_\mu$'s VEV~\cite{next+everett}.

The conclusion here is that any hypercharge carrying field next to the
string  may   fluctuate   and thus     lead  to   a source   term   in
Eq.~(\ref{Amueq}) which in turn is used in the Higgs field equations
\begin{equation} D_\mu D^\mu \Phi = -2{\delta V\over \delta |\Phi |^2}
  \Phi ,\ \  \ D_\mu   D^\mu h =  -2{\delta   V\over \delta |h|^2}   h
  ,\label{higgseq}   \end{equation}
through the  covariant derivatives.  It  turns out~\cite{next+everett}
that the  only  necessary   condition    for  this charging-up      to
spontaneously occur is   just  that the  corresponding  hypercharge be
massless.  This is where the various energy  scales are involved since
many cases are possible:  the first is that  the charged vector bosons
mixing the quarks and the leptons  at the GUT  scale acquire a mass at
or right after the string-forming symmetry breaking. In this case, the
massless  hypercharge is simply  the  electric charge and the  current
amplitude  is of the order $m_\star  \sim m  \sim \eta_{_{GUT}}$.  The
same applies also for the $W^\pm$ bosons of the electroweak sector (in
this case  $h$ is identified   with $H$).  These  bosons are  known to
exist, at an  energy scale $m_\star \sim  \eta _{_{EW}}$ that must  be
less than  $m$, since  we have not  observed any  string-forming phase
transitions  yet.  Thus, cosmic strings,  if they exist, must at least
carry an electroweak current.   Finally, there is the possibility that
another  phase transition occurs at  an intermediate  energy scale, in
which case the  corresponding currents  might  fairly well be  charged
coupled  or neutral because  it is sufficient to  build up the current
that the hypercharge  be massless only over  a finite region of space,
but not necessarily over all of it.  Thus, seen from the outside, such
a current-carrying string would be of the neutral kind~\cite{neutral}.

\section{Conclusions}

By   putting together previous  results   on the internal structure of
cosmic  strings, we have  shown that  they become current-carrying, at
least   at the electroweak scale.     This is because the  electroweak
symmetry  is restored around  the  string~\cite{symres}, giving rise to
spontaneous current generation due to charged vector bosons condensing
in the   string~\cite{scg,next+everett}.    If there   are appropriate
intermediate scales in  the   theory,  then analogous   results  hold.
However, this  is not the only  way a string might acquire spontaneous
gauge vector  current. Charged bosons coupling  quarks and leptons are
also  perfectly reasonable candidates  for that purpose.  In fact, it
entirely depends  on  the  actual  scale  at which  the string-forming
symmetry breaking occurs.

One cosmological consequence needs to  be emphasized at this point: in
the  case where $m\sim m_\star  \sim \eta_{_{GUT}}$,  one ends up with
current-carrying  cosmic strings whose   internal structure, i.e.  the
equation of state relating the energy per unit length and the tension,
looks  much like~\cite{scg}  that  derivable  for the  charged coupled
Witten      model~\cite{witten,supcond}.  But   using         Carter's
formalism~\cite{formal} to  evaluate   the stability~\cite{genstab} of
vorton remnants, it was found~\cite{witstab} that  such an equation of
state always leads to some stable  states.  Since the current-carrying
state represents  a  minimum of   the total  energy, they are  quantum
mechanically stable   as well. Thus  such  strings are   ruled out and
necessarily constrain the corresponding particle physics models.

It had  previously been shown~\cite{vortons,magnet} that the existence
of  vortons as stable  remnants  of  superconducting strings  severely
constrains the  underlying particle theory.  In Ref.~\cite{magnet}, it
was  also shown  that for  the  universe to be radiation  dominated at
nucleosynthesis,   GUT-scale   strings must    not  produce vortons at
temperatures above $\sim  10^6$~GeV.  Similarly for intermediate-scale
strings  becoming superconducting at the  same phase transition (i.e.,
with  $m\sim  m_\star$)  then  that   scale $m$   must be  less   than
$10^{10}$~GeV~\cite{magnet}  whilst a less conservative estimate gives
this to be  an upper bound of $m  \alt 10^4$~GeV only (the articles by
B.   Carter  in Refs.~\cite{vortons}).    However, our generic results
have shown  that cosmic     strings  become superconducting  at    the
electroweak scale.  In this case, the constraints are rather different
and two regimes need to be considered.   If strings are in the scaling
regime by the time of the electroweak phase  transition, then they are
sufficiently dilute ($\sim 1$ per horizon volume) that vorton remnants
are harmless.  If however, they  are in the friction dominated regime,
and hence copious, then the above constraints apply.  Strings produced
at  scales $m\agt 10^{10}$~GeV  are   in  the  scaling regime  by  the
electroweak   phase   transition.  Cosmological  consequences   of GUT
strings becoming  superconducting at  the electroweak phase transition
are currently   under   investigation  and  will be   discussed   in a
forthcoming publication.

\section*{Acknowledgements}

We thank B. Carter, T. W. B. Kibble and T. Vachaspati for discussions.
We acknowledge PPARC and a British Council-Alliance exchange grant for
financial support, and CERN for hospitality.

\end{document}